\documentclass[floatfix,twocolumn,aps,prc]{revtex4-2}

\usepackage{color,amsmath,amssymb,epsfig,graphicx}
\usepackage{bm}
\usepackage{mathptmx, courier, pifont}
\usepackage[scaled=0.92]{helvet}
\usepackage[T1]{fontenc}
\usepackage{textcomp}
\usepackage[colorlinks=true,linkcolor=blue,filecolor=blue,urlcolor=blue,citecolor=blue]{hyperref}

\usepackage{multirow}
\usepackage{longtable}
\usepackage[english]{babel}
\usepackage[autostyle, english = american]{csquotes}
\MakeOuterQuote{"}

\begin{document}
\author{Sameer Ahmad Mir}
\email{sameerphst@gmail.com}
\affiliation{Department of Physics, Jamia Millia Islamia, New Delhi, 110025, India}
\author{Nasir Ahmad Rather}
\affiliation{Department of Physics, Jamia Millia Islamia, New Delhi, 110025, India}
\author{Iqbal Mohi Ud Din}
\affiliation{Department of Physics, Jamia Millia Islamia, New Delhi, 110025, India}
\author{Saeed Uddin}
\email{suddin@jmi.ac.in}
\affiliation{Department of Physics, Jamia Millia Islamia, New Delhi, 110025, India}

\title{Hadron Production in Ultra-relativistic Nuclear Collisions and Finite Baryon-Size Effects}

\begin{abstract}
\noindent
We investigate relative hadron yield production of various like and unlike mass particles in ultra-relativistic heavy ion collisions by employing a statistical thermal model with finite-sized baryons (antibaryons) to imitate the hard-core repulsive interactions leading to the excluded volume type effect. A strong evidence of strangeness suppression relative to the non-strange ones, mainly pions, particularly at higher energies is also observed. This study also indicates that at chemical freeze-out the particle ratios and strangeness suppression in the system obtained theoretically are sensitive to baryonic (antibaryonic) hard-core radius ($r_B$). A comparison with earlier analysis involving the strangeness suppression effect is made where baryons and antibaryons were treated as point-like particles. The available experimental data showing energy dependence of various particle ratios are well described throughout the range of centre-of-mass energy ($\sqrt{s_{NN}}$). The value of hard-core radius between 0.76 to 0.79 fm is found to fit the data quite well using $\chi^{2}$ minimization technique. Two different chemical freeze-out stages are found where the earlier one belongs to baryonic (hyperonic), antibaryonic (antihyperonic) states and the later one to mesonic degrees of freedom.
\end{abstract}

\date{\today}
\maketitle
\section{Introduction}
\noindent
\label{intro}
One of the principal objectives of the ultra-relativistic heavy ion collisions (URHIC) has been the creation of a hot and dense hadronic matter, also called fireball, leading to the production of various hadron species. This process holds a distinct advantage over~\textit{p-p} collisions as it induces multiple secondary inelastic and elastic collisions among the particles produced within the fireball~\cite{hp1,hp3,hp4,hp2,hp5}.
According to quantum chromodynamics (QCD), a phase consisting of deconfined quarks and gluons at very high temperature $(T)$ and/or densities or baryon chemical potential (BCP) is expected to exist while at low temperatures and/or densities, the quarks and gluons remain confined within hadrons and the system is described by a phase consisting of hadronic degrees of freedom. Hence emphasis has been laid to explore the QCD phase diagram in $T-\mu_{B}$ plane~\cite{hp6,hp7,hp8,hp9,hp10,hp11,hp12,hp13}. The BCP $(\mu_{B})$ plays a pivotal role reflecting the abundance of baryons over antibaryons and serves as a crucial statistical thermodynamic parameter controlling the net baryon content of the hadronic fireball system.
The highest RHIC energy and the subsequent LHC-ALICE experiments have successfully created hadronic matter with significant excitation, leading to the formation of a considerably larger system in URHIC. The resulting matter from these collisions is nearly symmetric in baryons and antibaryons \cite{hp14} which is a consequence of the substantial nuclear transparency observed in URHIC at these energies. As a result, such a system maintains a small BCP ($\sim$1 MeV), indicating a minimal excess of baryons over antibaryons. In contrast, hadronic matter created in experiments like BNL{\textendash}AGS, CERN{\textendash}SPS, and lower-energy RHIC experiments  ($\sqrt{s_{NN}}\:\sim$ 4.7 MeV {\textendash} 11.6 GeV) retains a larger BCP ($\sim$500 {\textendash} 250 MeV) due to a significant excess of baryons over antibaryons, attributed to a high degree of nuclear stopping power effect during URHIC in these experiments. The typical temperature of the formed matter is within the range of 110 {\textendash} 160 MeV~\cite{hp15,hp16,hp17,hp18,hp19,hp20,hp21,hp22}.

In the very initial stages of URHIC, there  are observable indications of a phase exhibiting dense partonic degrees of freedom~\cite{hp6,hp7,hp8,hp9,hp10,hp22,hp23,hp24,hp25,hp26,hp27} later transitioning into one characterized by hadronic degrees of freedom~\cite{hp28,hp29,hp30,hp31}. In these collisions the hot and highly dense secondary partonic matter, which is composed of quarks and gluons in the fireball is initially in a state of pre-equilibrium where the constituent particles (partons) undergo multiple elastic and inelastic collisions. These interactions result in a large particle densities leading to an enormously high pressure within the system. Consequently, the fireball undergoes an expansion both due to multi-particle production and a hydrodynamic flow indicating a collective effect leading to an increase in the system size. The system eventually attains a state of sufficiently high thermo-chemical equilibrium with a substantially large number of particles, enabling the application of statistical thermal models. In case of sufficiently high initial temperature ($T$) and net baryon density ($\rho$), the system may be in a state called quark-gluon plasma (QGP)~\cite{hp32,hp33,hp34,hp35}. As the system further expands then at a critical stage a first-order phase transition may occur, resulting in a mixed phase consisting of QGP and a phase typically described by a hadron resonance gas (HRG) system. The particles (i.e. hadrons) in the HRG phase continue to interact even after the completion of the first order phase transition resulting in further production of various hadrons through the collisions of various hadronic species in this phase. This can lead to a reasonably high degree of thermal and chemical equilibration of the hadronic species in the HRG. The HRG continues to expand, grow in size and cool. This dynamic evolution of the HRG system is followed by a chemical freeze-out. This situation occurs when the mean free paths of the hadrons approach a comparable length scale with the overall size of the system. The interactions between hadrons almost diminish at freeze-out, leading to a state where hadron abundance is frozen in time~\cite{hp36,hp37,hp38,hp39}. 
The study of the final state relative particle yields following the freeze-out can serve as a crucial tool for studying the characteristics of the hadronic matter and hence exploring its complex properties~\cite{hp40,hp41,hp42,hp43,hp43b,hp44,hp46,hp47,hp48}.

It has been further demonstrated earlier that considering hadrons as point-like particles, fails to accurately reproduce the fundamental properties of nuclear matter in its ground state. Moreover, it is also notable that within the confines of a thermal model, a first-order QGP-HRG phase transition cannot be established properly with a sufficiently large number of hadronic degrees of freedom taken into account in the HRG phase~\cite{hp50,hp51,hp52} if hadrons are considered to have almost vanishing sizes. This phenomenon primarily occurs because in absence of any repulsive hard-core interaction among the hadrons within a given physical volume of the system, a very large number of point-like hadronic resonances can be thermally excited at high temperatures. It is found that within the framework of a statistical thermal model and a first-order quark-hadron phase transition, the pressure in the HRG phase with sufficiently large number of point-like hadronic degrees of freedom becomes more than the pressure in the QGP phase (i.e. $P_{HRG}$ > $P_{QGP}$) and consequently the system undergoes a transition back to the HRG phase at high temperatures. The consideration of this aspect thus becomes essential for understanding the phase transition from the QGP phase to the HRG phase and its thermodynamic properties~\cite{hp5,hp53,hp54,hp55,hp56,hp57}.

It has been shown that this problem can be resolved by considering a hard-core repulsive interaction among hadrons. In the framework of statistical thermal models, the hard-core repulsion which essentially exists between a pairs of baryons and a pairs of antibaryons can be taken into account in a somewhat simple way by assigning them a finite size in the HRG phase. Under this condition a given baryon in the system cannot move freely in the entire volume of the system but only in the available volume which is free from other baryons. Hence an effective proper volume $\nu$ is assigned to every baryonic (antibaryonic) specie. Following a van der Waals type treatment, the volume~\textit{V} is replaced by $V- \nu N$~\cite{hp58}. Though in strict sense the procedure is thermodynamically inconsistent, nevertheless it still provides reasonably good results including the ground state properties of the cold matter. The repulsive force is assumed to act between a pair of baryons or a pair of antibaryons \cite{hp55,hp59}.
The analysis of hadronic equation of state so far have been employed within the framework of statistical models through various computational tools like THERMUS \cite{thermus}, Thermal-FIST~\cite{fist}, and THERMINATOR~\cite{therminator}. These authors have calculated several thermodynamic parameters of the system. However, no analysis of the energy dependence of various particle ratios over a wide range of energy has been done.

It is therefore worth investigating that to what extent such hard-core repulsive interactions can affect the relative hadronic ratios in the hot and dense hadronic medium formed at various collision energies $(\sqrt{s_{NN}})$ in URHIC. In some recent works~\cite{refree1,refree2, hpref12} analysis of hadronic ratios have been carried out. However these authors have not taken into account the hadronic hard-core repulsive interactions and hence have treated hadrons as point-like particles. In this work, we have analyzed the available experimental data on various particle ratios where we have treated the quantity $\nu$ as a free parameter of the system in addition of its temperature $(T)$ and BCP ($\mu_{B}$) to investigate their effects over a wide range of collision energies $(\sqrt{s_{NN}})$ in URHIC by making use of a statistical thermal model incorporating the finite hadron sizes by considering their effective proper volumes.

The manuscript is structured as follows. In section \ref{statisticalappraoch}, we briefly describe the hadron resonance gas model. 
In section \ref{results}, we delve  into a comprehensive discussion of the results obtained. Finally we summarise and conclude our results in section \ref{summary}.

\section{The Statistical Approach}
\label{statisticalappraoch}
\noindent
For point-like (approximately ideal) non-interacting particles, whose number is not conserved in the system, thermodynamic quantities can be derived by defining the partition function of the grand canonical ensemble (GCE). For such an ensemble at a specific temperature and chemical potential, the partition function can be expressed as

\begin{equation}
    Z^{ni}=Tr\left[exp\left( \frac{-1}{T_{i}}(\hat{H}-\mu_{i}\hat{N})\right)\right]
    \label{eqn:partfunc}
\end{equation}
Where $\hat{H}$ and $\hat{N}$ are the Hamiltonian and particle number operators, respectively. The superscript $ni$ stands for non-interacting particles. The $T_{i}$ and $\mu_{i}$ are the temperature and chemical potential of the $i^{th}$ hadronic specie in the multi-component hadronic resonance gas system.

As discussed above the hadrons at freeze-out are nearly free hence for such a case the trace in the above equation can be performed by using the particle number representations for fermions and bosons \cite{hp60}. Further in the continuum limit this can be reduced to an integral form to obtain the logarithm of the grand canonical partition function for particles as \cite{hp61,hp62}

\begin{equation}
    lnZ^{ni}(T,\mu,V)=\frac{gV}{2\pi^{2}}\int{k^{2}ln\left[ 1+ \eta e^{\left({\frac{-( E+\mu_{i})}{T_{i}}}\right)} \right]}dk
    \label{eqn:grandpartfunc}
\end{equation}
where $\eta=$ +1 corresponds to fermions while -1 corresponds to bosons. The above equation can be reduced to the following form

\begin{multline}
    lnZ^{ni}(T,\mu,V)=\frac{gV}{6\pi^{2}T_{i}}\int{\frac{dk k^{4}}{\sqrt{k^{2}+m^{2}}}} \times\\
    \left[ \frac{1}{e^{[\sqrt{k^{2}+m^{2}}-\mu_{i}]/T_{i}}+\eta}\right]
    \label{eqn:grandpartfuncfinal}
\end{multline}

The corresponding expression for the antiparticles can be obtained by replacing $\mu_{i}\rightarrow -\mu_{i}$~\cite{hp12,hp13,hp60,hp63,hp64,hp65}. The integral is over the momentum of the single particle having rest mass \textit{m}. The ensemble averaged (mean) number density of the point-like non-interacting $i^{th}$ hadronic specie ($n_{i}^{ni}$) can be obtained in a thermodynamically consistent manner by taking the derivative

\begin{equation}
    n_{i}^{ni}= \left( \frac{\partial ln Z_{i}^{ni}}{\partial\mu_{i}}\right)_{T,V}
    \label{eqn:numden1}
\end{equation}

Using Eq.'s~\eqref{eqn:grandpartfuncfinal} and~\eqref{eqn:numden1}, the number density can be obtained by using Boltzmann approximation (for simplicity) as

\begin{equation}
    n_{i}^{ni}=\frac{g_{i}}{2\pi^{2}}T^{3}\lambda_{i}W(m_{i}/T)
    \label{eqn:numden}
\end{equation}

Where $m_{i}$ and $g_{i}$ are the mass and the spin-isospin degeneracy factor of the $i^{th}$ type of hadronic specie. The \textit{T} is the thermal temperature of the system and $W (m_{i} /T)= (m_{i}/T)^{2} K_{2}(m_{i}/T)$ with $K_{2}$ as the modified Bessel function. The quantity $\lambda_{i}=e^{\mu_{i}/T}$ is the fugacity of the $i^{th}$ type of hadronic specie.

The effect of the hard-core repulsive interaction among the baryons, resulting in a finite size or excluded volume type effect, can be taken into account. In a grand canonical ensemble system with \textit{N} mean number of particles, this procedure corresponds to a substitution of the system volume \textit{V} by the available mean volume $V- \nu N$, where $\nu$ is the particle’s proper volume. This proposal is further supported by calculations within statistical mechanics for a system of hard non-deformable spheres of baryon radius $r_{B}$. In this case, $\nu$ is replaced by $b$, which equals the particle hard-core volume $4\pi r^{3}_{B}/3$ multiplied by a factor of 4~\cite{hp66}. One can write the modified expression for the mean number density ($n_{i}^{fs}$) of any given $i^{th}$ finite-size baryon specie as

\begin{equation}
    n_{i}^{fs}=n_{i}^{ni}\left(1+\sum_{j}\nu_{j}n_{j}^{ni}\right)^{-1}
    \label{eqn:finitenumden}
\end{equation}

The summation over the index $j$ (which also includes $i$) is performed over all baryonic species to incorporate the finite-size effect of all baryons. One can obtain a similar expression for the antibaryons.
The $\nu_{j}$ is the proper volume of the $j^{th}$ baryonic (antibaryonic) specie. The mesons are assumed to be free of such hard-core repulsion \cite{hp55,hp57}. The result in Eq.~\eqref{eqn:finitenumden}, though thermodynamically inconsistent, can still provide a good estimate of the $relative$ number densities in the HRG model.

In order to obtain a comparison between theoretical results and experimental data, we need to have a reasonably good estimate of the values of temperature and BCP at the chemical freeze-out at different collision energies, i.e., $\sqrt{s_{NN}}$. A convenient way to achieve this is to establish a relationship between the freeze-out parameters, i.e., temperature $T$, BCP $\mu_{B}$, and $\sqrt{s_{NN}}$. The following ansatz has been employed to fit the extracted values of $T$ and $\mu_{B}$ from the experimental hadronic yields and their spectra~\cite{hp50,hp67,hp68,hp69,hp70,hp71}.

\begin{equation}
    T=c-d\mu^{2}_{B}-e\mu^{4}_{B}
    \label{eqn:temp}
\end{equation}

\begin{equation}
    \mu_{B}=\frac{a}{1+b\sqrt{s_{NN}}}
    \label{eqn:chempot}
\end{equation}

This kind of ansatz has been used effectively to study the properties of the hot and dense medium formed in URHIC within thermal models at different collision energies. Analysis over a broad energy range~\cite{hp64,hp72,hp12} have significantly contributed to the ongoing efforts to understand the chemical freeze-out criteria~\cite{hp11,hp67,hp75,hp76,hp77,hp79,hp79b,hp80}.

\section{Results and Discussion}
\label{results}
\noindent
The Eq.~\eqref{eqn:numden} provides us with the ensemble averaged number density of the $i^{th}$ type of hadron in the system. The chemical potential of the $i^{th}$ hadronic specie is defined as~\cite{hp60,hp82,hp83,hp84,hp86,hp87,hp88,hp90}

\begin{equation}
    \mu_{i}=N_{q}\mu_{q}+N_{s}\mu_{s}
    \label{eqn:chempot1}
\end{equation}

Where $N_{q}$ and $N_{s}$ are the number of valence light (u,d) and strange (s) quarks, respectively in the $i^{th}$ type of hadronic specie. The quantity $\mu_{q}=\mu_{B}/3$ is the light quark chemical potential, where $\mu_{B}$ is the BCP of the system and $\mu_{s}$ is the strange chemical potential (SCP) which essentially controls the net strangeness content of the system.
This is sufficient to define the fugacity of all the hadronic species i.e., Kaon fugacity is given by $\lambda_{K}=\lambda_{q}\lambda^{-1}_{s}$,  antikaon fugacity as $\lambda_{q}^{-1}\lambda_{s}$, non-strange baryon fugacity is $\lambda_{B}=\lambda^{3}_{q}$, singly strange hyperon ($\Lambda, \Sigma$) fugacity as $\lambda_{\Lambda,\Sigma}=\lambda^{2}_{q}\lambda_{s}$, doubly strange hyperon ($\Xi$) fugacity as $\lambda_{\Xi}=\lambda_{q}\lambda^{2}_{s}$ and triply strange hyperon i.e., ($\Omega$) fugacity is $\lambda_{\Omega}=\lambda^{3}_{s}$ etc. For all antibaryons (antihyperons), their fugacities will be the inverse of their corresponding baryon’s (hyperon’s) fugacities. The value of $\mu_{s}$ is fixed by using the criteria of overall strangeness conservation, which requires that the total strangeness content of the system must be equal to its total antistrange content. This yields the following equation involving number densities of various strange hadronic species included in our system

\begin{multline}
    \sum_{m}n_{\overline{K}{_{m}}}+\sum_{\alpha}n_{\alpha}+2\sum_{\beta}n_{\beta}+3\sum_{\delta}n_{\delta} =\\
    \sum_{m}n_{{K}{_{m}}}+\sum_{\alpha}n_{\overline{\alpha}}+2\sum_{\beta}n_{\overline{\beta}}+3\sum_{\delta}n_{\overline{\delta}}
    \label{eqn:strconser}
\end{multline}

In the above equation, the index $m$ represents kaonic (antikaonic) degrees of freedom, while the indices $\alpha$, $\beta$ and $\delta$ represent singly, doubly, and triply strange (antistrange) hyperons, respectively.
The factor 2 with doubly strange (antistrange) terms is for cascades (anticascades) resonances. Similarly, factor 3 takes into account the triply strange (antistrange) nature of omega (antiomega) baryon.

\begin{figure}[b]
    \hspace{0.0cm}\includegraphics[scale=0.55]{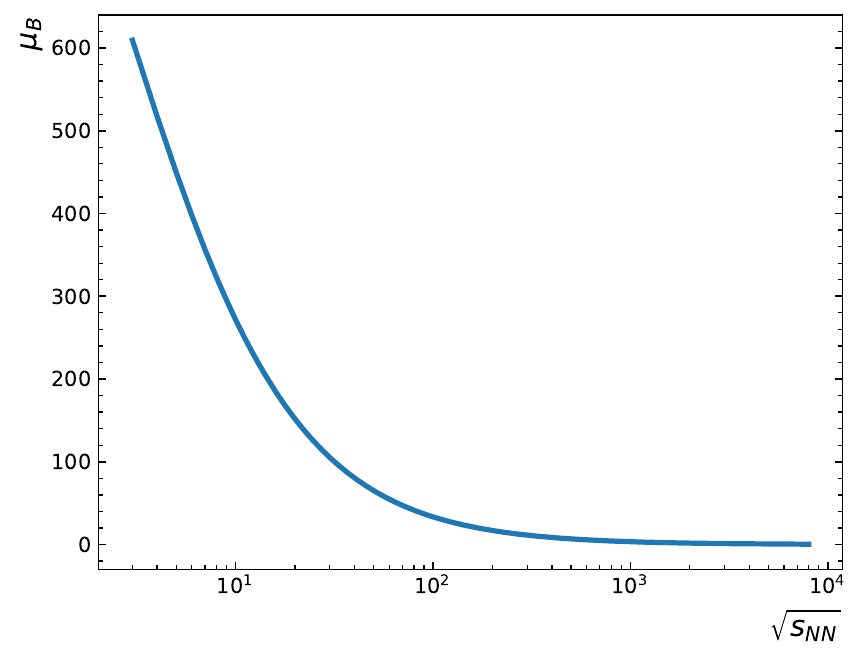}
    \hspace{0.0cm}\includegraphics[scale=0.55]{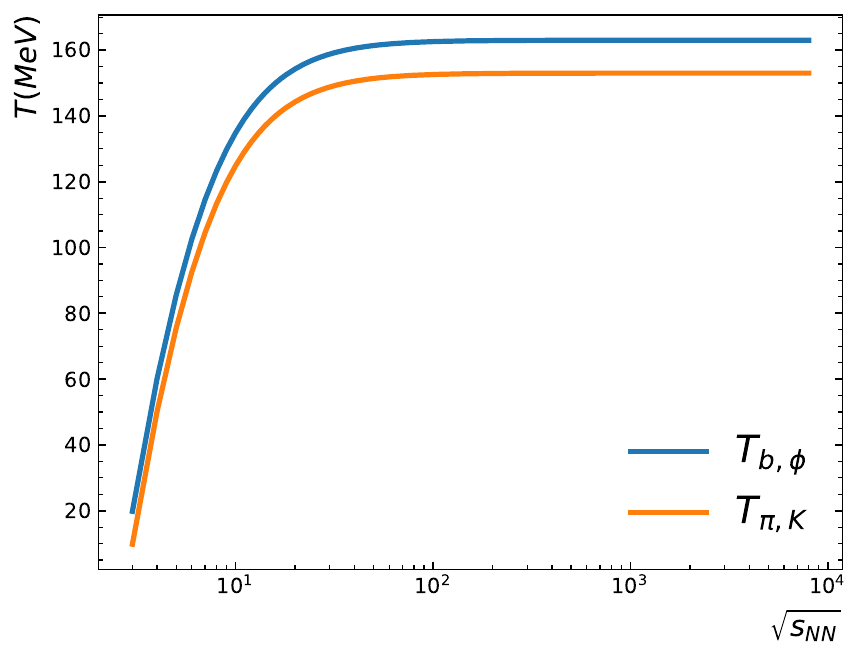}
    \caption{Dependence of baryon chemical potential ($\mu_{B}$) and temperature (T) on $\sqrt{s_{NN}}$ for the two freeze-out scenarios}
    \label{fig:cpottemp}
\end{figure}

In the present work we have analyzed the dependence of several particle ratios on the collision energy in URHIC i.e. ($\sqrt{s_{NN}}$). The particle ratios obtained by using the thermal model approach are fitted on the experimental data obtained at different collision energies. The temperatures and chemical potentials for different collision energies are fixed through Eq.'s~\eqref{eqn:temp} and~\eqref{eqn:chempot}. Initially the baryonic (antibaryonic) hard-core radius used in Eq. \eqref{eqn:finitenumden} and the parameters occurring in Eq.'s~\eqref{eqn:temp} and~\eqref{eqn:chempot} are varied to obtain the best fit for the experimental $\overline{p}/p$ ratio as a function of $\sqrt{s_{NN}}$ over a wide range. The minimum chi-squared is used as the best fit criteria for the goodness of fit, defined as

\begin{equation}
    \chi^{2} = \sum_{i=1}^{N}\frac{(R_{i}^{th}-R_{i}^{exp})^{2}}{\sigma_{i}^{2}}
    \label{eqn:chisq}
\end{equation}

The $R_{i}^{th}$ and $R_{i}^{exp}$, respectively, are the theoretical and experimental values of the particle ratios for the $i^{th}$ value of collision energy. The $\sigma_{i}$ represents the error in the experimental measurements of the ratios for corresponding energy.

\begin{figure}[h]
    \hspace{0.0cm}\includegraphics[scale=0.55]{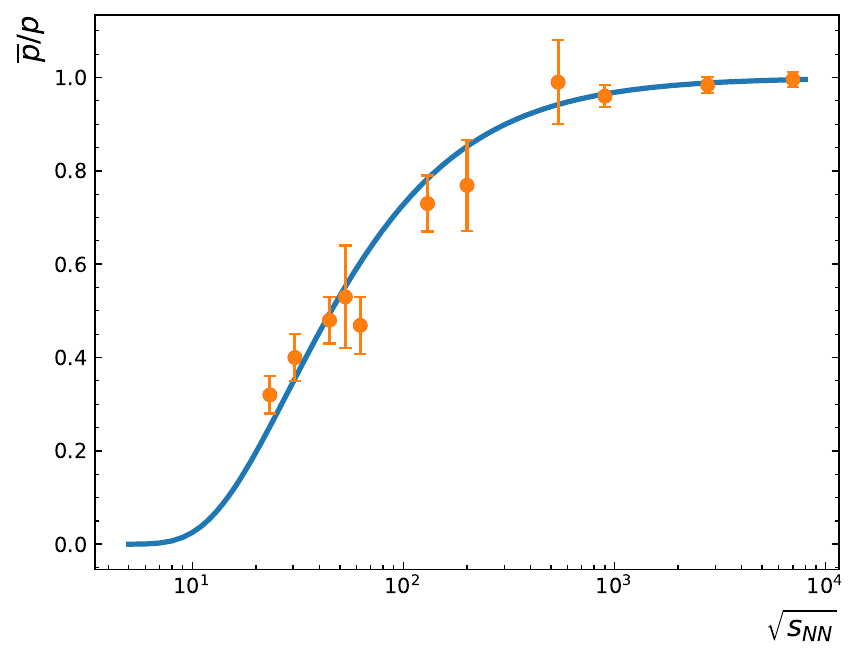}
    \caption{$\overline{p}/p$ dependence on $\sqrt{s_{NN}}$}
    \label{fig:protons}
\end{figure}

In Fig.~\ref{fig:cpottemp}, we have first shown the dependence of the BCP and temperature on the centre of mass collision energy ($\sqrt{s_{NN}}$) defined by Eq.'s~\eqref{eqn:temp} and \eqref{eqn:chempot} for the best fitted values of the parameters obtained through the minimum $\chi^{2}$ fit procedure of the $\overline{p}/p$ ratio using the experimental data of RHIC-STAR and ALICE-LHC. This is essentially done as protons (antiprotons) are among the most abundantly produced baryons (antibaryons) in nuclear collisions hence their abundance is used here to fix the values of BCP and temperature for all other baryons as well.
In Fig.~\ref{fig:protons} the best fitted curve is shown by the solid line obtained by minimizing the $\chi^{2}$/dof value. We have used the best fitted curve of the $\overline{p}/p$ data set~\cite{hpref1,hpref2,hpref3,hpref4,hpref5,hpref6,hpref12} to fix the parameters of the ansatz given by Eq.'s~\eqref{eqn:temp} and~\eqref{eqn:chempot}, along with the baryon (antibaryon) hard-core radius of protons (antiprotons). For simplicity we have chosen the hard-core radius to be same for all baryons and antibaryons for each case separately i.e., $\overline{\Lambda}/\Lambda$, $\overline{\Xi}/\Xi$, and $\overline{\Omega}/\Omega$ ratios.

\begin{figure}
    \includegraphics[scale=0.55]{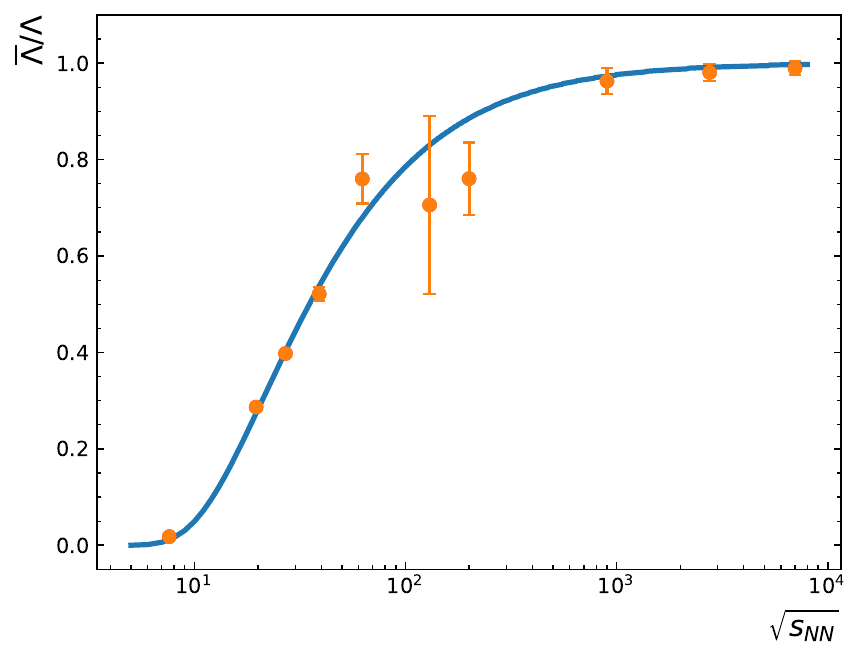}
    \caption{$\overline{\Lambda}/\Lambda$ dependence on $\sqrt{s_{NN}}$}
    \label{fig:lambdas}
\end{figure}

\begin{figure}
    \includegraphics[scale=0.55]{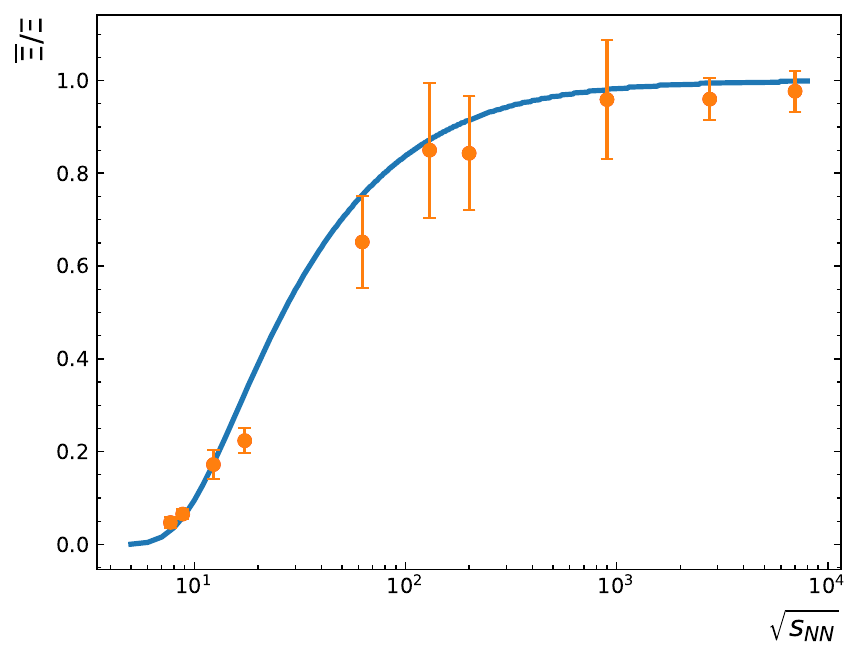}
    \caption{$\overline{\Xi}/\Xi$ dependence on $\sqrt{s_{NN}}$}
    \label{fig:cascade}
\end{figure}

\begin{figure}
    \includegraphics[scale=0.55]{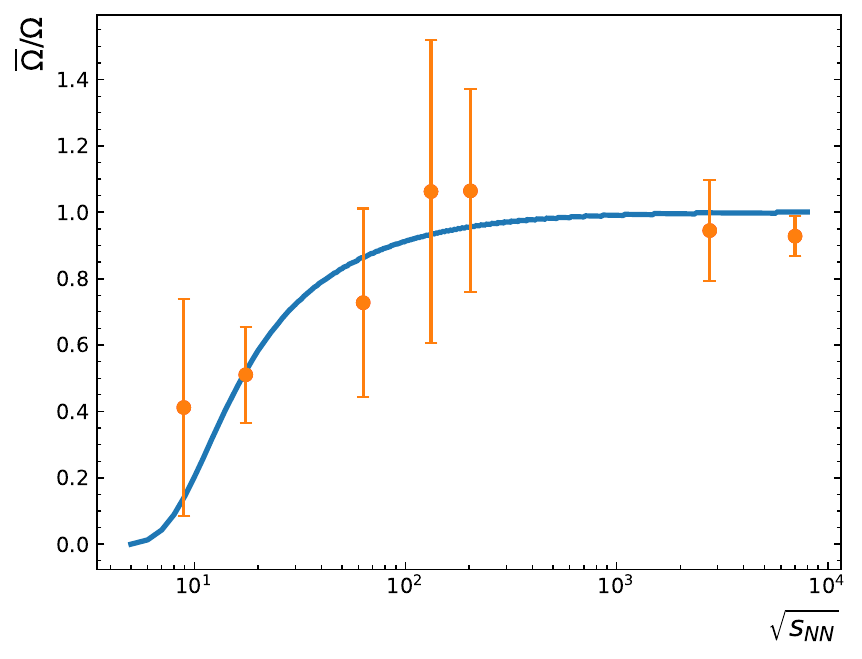}
    \caption{$\overline{\Omega}/\Omega$ dependence on $\sqrt{s_{NN}}$}
    \label{fig:omega}
\end{figure}

The experimental data and their respective theoretical thermal model fits for the cases of $\overline{\Lambda}/\Lambda$, $\overline{\Xi}/\Xi$, and $\overline{\Omega}/\Omega$ \cite{hpref2,hpref4,hpref7,hpref10,hpref11} are shown in Figs.~\ref{fig:lambdas},~\ref{fig:cascade}, and~\ref{fig:omega} respectively. Contribution of weakly decaying hadrons upto 2 GeV mass after freeze-out is also taken into account \cite{hp2GeV1,hp2GeV2}.

We find that all four antibaryon to baryon ratios increase with increasing center-of-mass energy ($\sqrt{s_{NN}}$) up to 200 GeV, then start to saturate beyond 200 GeV approaching a limiting value $\sim$1. This indicates that there is a significant baryon antibaryon asymmetry in the region well below $\sqrt{s_{NN}}$ $\sim$ 200 GeV, and the system becomes approximately symmetric towards higher energies, and consequently, the antibaryon to baryon ratios approach unity.

\begin{figure}
    \centering
    \includegraphics[scale=0.55]{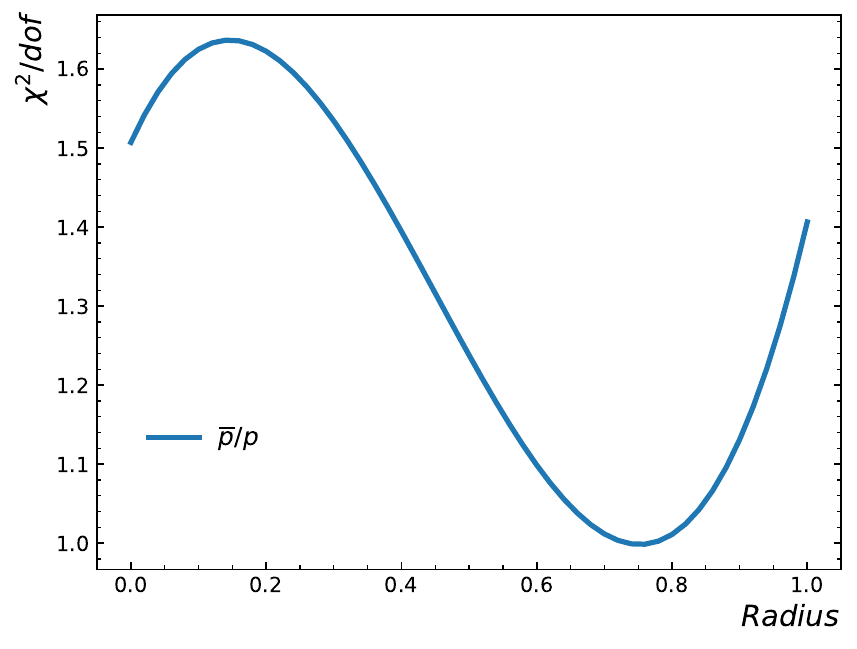}
    \caption{$\chi^{2}$ per degree of freedom varying with increasing radius}
    \label{fig:chisquare}
\end{figure}

In the framework of our thermal model we find that the BCP decreases from a value of around 600 MeV (for $\sqrt{s_{NN}}$ $\sim$ 5 GeV) to approximately 20 MeV (for $\sqrt{s_{NN}}$ $\sim$ 200 GeV) and asymptotically approaches zero value towards the LHC energies. On the other hand the freeze-out thermal temperature increases with increasing centre-of-mass energy and also tends to saturate at approximately 163 MeV for $\sqrt{s_{NN}}$ beyond 200 GeV for baryons (antibaryons). 
The decrease in the BCP with increasing $\sqrt{s_{NN}}$ occurs due to an almost equally high degree of excitation in the baryonic as well as antibaryonic sectors leading to an almost baryon symmetric hot hadronic matter. This phenomenon is also found to be related to the occurrence of the nuclear transparency effect in the URHIC at the highest RHIC energies and beyond~\cite{hp36}, leading to a saturation in temperature of the system formed under such a condition and significant decrease in the BCP ($\mu_{B}$). The above described scenario is reflected in the ansatz used in Eq.'s~\eqref{eqn:temp} and~\eqref{eqn:chempot}. As already discussed, we have treated the baryon (antibaryon) hard-core radii ($r_{B}$) as one of the free parameters in our calculation. The values of the best fitted hard-core radii for the cases shown in Fig's~\ref{fig:protons},~\ref{fig:lambdas},~\ref{fig:cascade}, and~\ref{fig:omega} are 0.78 fm, 0.76 fm, 0.79 fm and 0.78 fm, respectively. We thus find that the best fitted values of $r_{B}$ for the four cases are very close and lie in the range 0.76 - 0.79 fm. 
The radius assigned to all baryons in the works mentioned in~\cite{munzinger:1999pkr,hp70} is 0.80 fm which lies close to the radius obtained in the present work.
The minimum $\chi^{2}$/dof values for the four cases are 0.95, 1.34, 1.89 and 0.38, respectively. 
For the purpose of illustration, we have shown in Fig.~\ref{fig:chisquare} the variation of $\chi^{2}$/dof values for $\overline{p}/p$ ratio with increasing hard-core radius, \textit{\lq r\rq}.

In our analysis, we have found an evidence of a double chemical freeze-out scenario corresponding to baryons and mesons. Choi and Lee~\cite{choilee} have also discussed the occurrence of two different freeze-outs for hadrons, where a chemical freeze-out occurs earlier at a higher temperature and the thermal freeze-out occurs later at a lower temperature.
In the present work it is seen that a lower chemical freeze-out temperature, as compared to the baryons (antibaryons) is required to explain the experimental data distribution over the entire range of energy considered for the mesonic (kaonic and pionic) degrees of freedom. The $\phi(s\bar{s})$ meson due its small reaction cross section with other hadrons is assumed to freeze-out earlier than $K$ and $\pi$ along with baryons and is also supported by earlier study of Flor et.al.~\cite{Flor}.

We have analyzed the available experimental $K^{-}/K^{+}$ data over a wide range of energy \cite{hpref3,hpref7,hpref13}. The best-fitted value of the parameter $c$ in Eq. \eqref{eqn:temp} is found to be 153 MeV in this case, while for the baryonic (antibaryonic) and $\phi$ sector it is 163 MeV. The value of the remaining parameters remain unchanged, which are; $a$ = 1.304 GeV, $b$ = 0.39 GeV$^{-1}$, $d$ = 0.38 GeV$^{-1}$, $e$ = 0.015 GeV$^{-3}$. In a recent work Waqas and Peng \cite{refree3} have shown that there is an evidence of more than one hadronic kinetic (thermal) freeze-out stages. They have analyzed the transverse momentum ($p_{T}$) distribution of different hadrons produced in URHIC.

\begin{figure}[h]
    \includegraphics[scale=0.55]{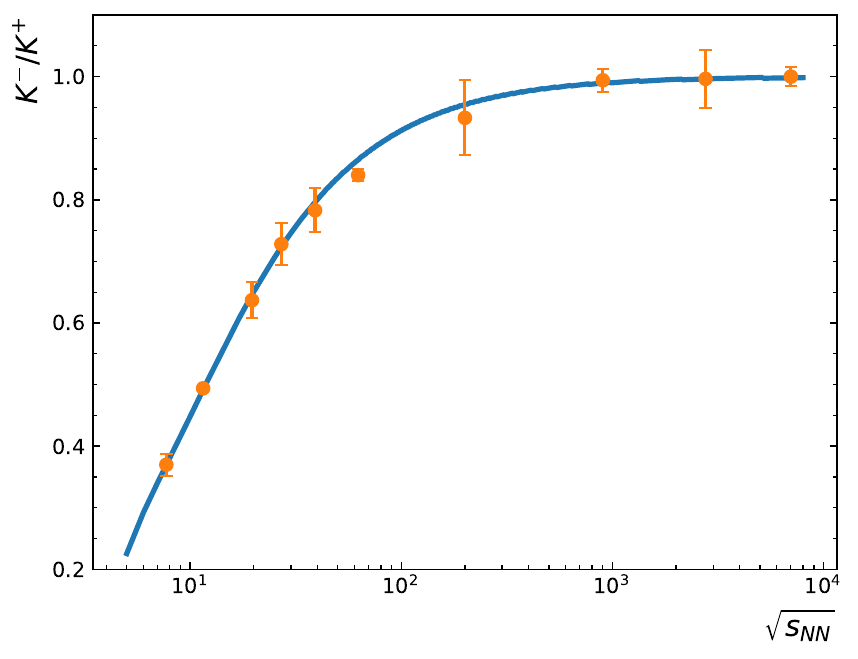}
    \caption{$K^{-}/K^{+}$ dependence on $\sqrt{s_{NN}}$}
    \label{fig:kaons}
\end{figure}

\begin{figure}
    \includegraphics[scale=0.55]{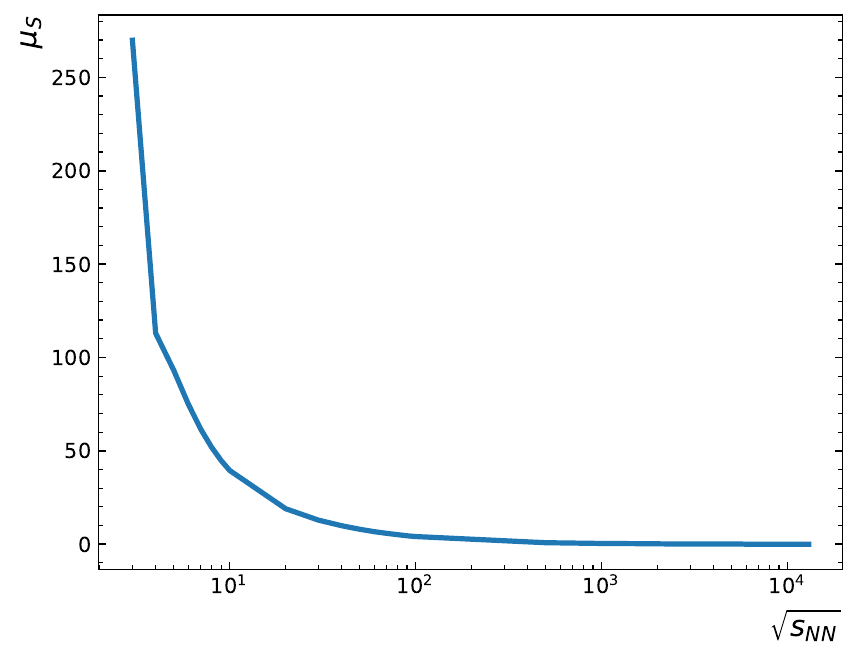}
    \caption{Extracted values of strange chemical potential ($\mu_{s}$)}
    \label{fig:cpotstrange}
\end{figure}

The increasing trend in the $K^{-}/K^{+}$ ratio with collision energy is again an indicator of baryon symmetric matter being produced at high energies. This can be further understood in light of the decreasing values of $\mu_{B}$ and $\mu_{s}$ with $\sqrt{s_{NN}}$ in Figs.~\ref{fig:cpottemp} and~\ref{fig:cpotstrange}, along with the increasing value of $T$ in Fig.~\ref{fig:cpottemp}. The reaction rates (involving nucleons) which populate the $K^{-}$ and $K^{+}$ phase spaces also indicate that in a baryon rich system the production rates of $K^{+}$ will be larger than those of the $K^{-}$. The reactions for the production of $K^{+}$ involving pion-nucleon and nucleon-nucleon interactions are mainly:

$\pi^{+}n\rightarrow K^{+}\Lambda$, $\pi^{0}p\rightarrow K^{+}\Lambda$, $pp\rightarrow p \Lambda K^{+}$, $np\rightarrow n \Lambda K^{+}$

While the $K^{-}$ producing reactions (involving antinucleons) are mainly:

$\pi^{-}\overline{n}\rightarrow K^{-}\overline{\Lambda}$, $\pi^{0}\overline{p}\rightarrow K^{-}\overline{\Lambda}$, $\overline{p}$ $\overline{p}\rightarrow \overline{p}\overline{\Lambda}K^{-}$, $\overline{n}$ $ \overline{p}\rightarrow \overline{n}\overline{\Lambda}K^{-}$.

Through the pion-pion reaction channel, the $K^{-}$ and $K^{+}$ are produced in equal proportion viz, $\pi \pi \rightarrow K^{+}K^{-}$. Thus the asymmetry in the abundance of $K^{-}$ and $K^{+}$ in the system is essentially due to difference in their production through the reaction channels involving nucleons and antinucleons.

Hence from above, it is evident that at lower values of $\sqrt{s_{NN}}$ where a baryon rich system is formed with a large excess of nucleons over antinucleons, the thermal production rate for $K^{+}$ will be larger than that of $K^{-}$, while at larger energies their production rates becomes almost equal.
The experimental data on the $K^{-}/K^{+}$ ratio clearly supports the thermal model predictions.

Another important relative hadronic yield is the kaon-to-pion ratio, as it provides a clear signature of the abundance of the light strange mesons (i.e., kaons) relative to the non-strange ones, especially the pions as they are copiously produced in the system and are an important indicator of the entropy content of the system~\cite{hpentropy1,hpentropy2}. With this aim we have analyzed the strange hadron abundance relative to the pion abundance in the system. We have obtained the $K^{-}/\pi^{-}$, $K^{+}/\pi^{+}$, and $\Xi^{+}/\pi^{-}$ ratios dependence on $\sqrt{s_{NN}}$. We have found that the theoretical results obtained in the framework of the thermal model significantly overestimate these experimental data at all energies~\cite{hpref3,hpref7,hpref9}. This supports the earlier idea of partial saturation of strange phase space~\cite{hp86,hppartial} relative to non-strange ones, mainly pions. We therefore define a strangeness suppression factor, $\gamma_{s}$, which accounts for the observed deviation from chemical equilibrium in strangeness sector. A value of $\gamma_{s}$ equal to unity would imply a chemical equilibration in strangeness sector. Several authors in the earlier works have discussed its significance and its dependence on energy and centrality in URHIC~\cite{hpref12,hp72,refree4,refree5,refree6,refree7,refree8,refree9}. We have in our case defined it as the mean of the ratios of the experimental values of these particle ratios to their respective theoretical ones. The third case i.e., $\Xi^{+}/\pi^{-}$ gives $\gamma_{s}^{2}$ (since $\Xi^{+}$ is a doubly strange hadron) hence we have to take this aspect into account in calculating the mean value of $\gamma_{s}$ as described above. This dependence of $\gamma_{s}$ on $\sqrt{s_{NN}}$ obtained by us is very well represented by an ansatz of the following type

\begin{equation}
    \gamma_{s}=\gamma_{s}^{o}+0.184 e^{-y(\sqrt{s_{NN}}-z)}
    \label{eqn:gammas}
\end{equation}

\begin{figure}
    \includegraphics[scale=0.55]{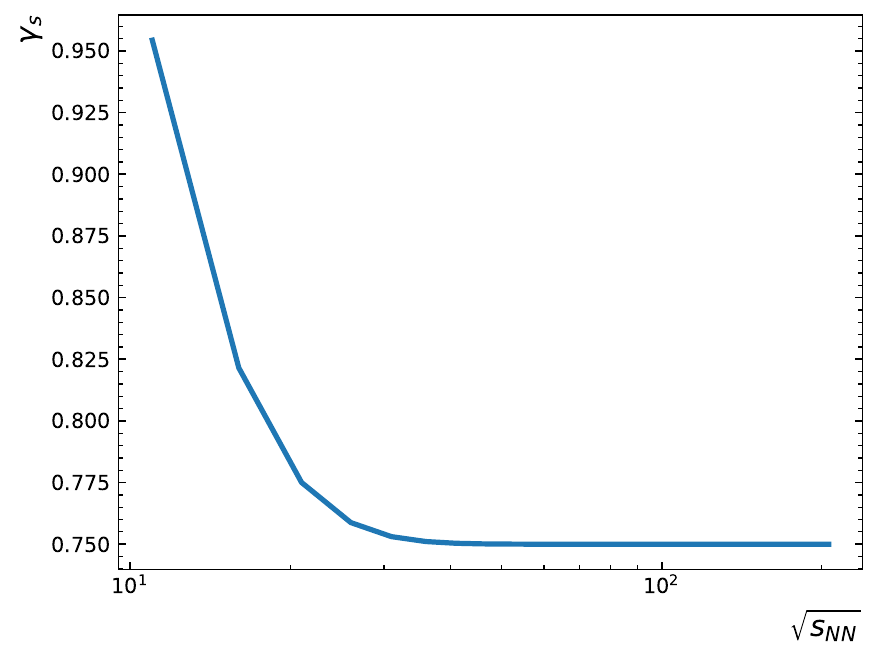}
    \caption{Dependence of strangeness suppression factor $\gamma_{s}$ on $\sqrt{s_{NN}}$}
    \label{fig:gammas}
\end{figure}

\begin{figure}
    \includegraphics[scale=0.55]{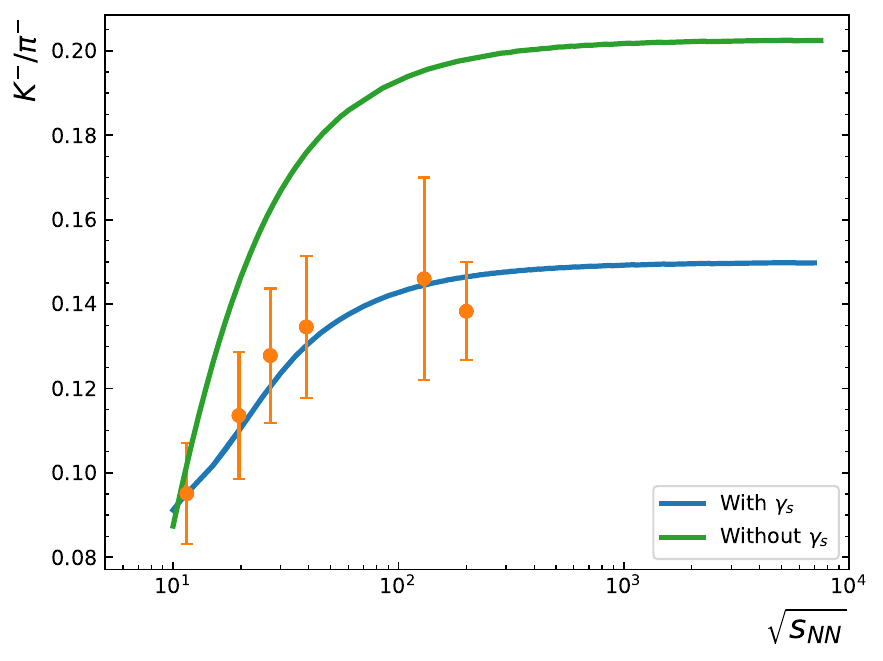}
    \caption{$K^{-}/\pi^{-}$ dependence on $\sqrt{s_{NN}}$}
    \label{fig:kaonpim}
\end{figure}

\begin{figure}
    \includegraphics[scale=0.55]{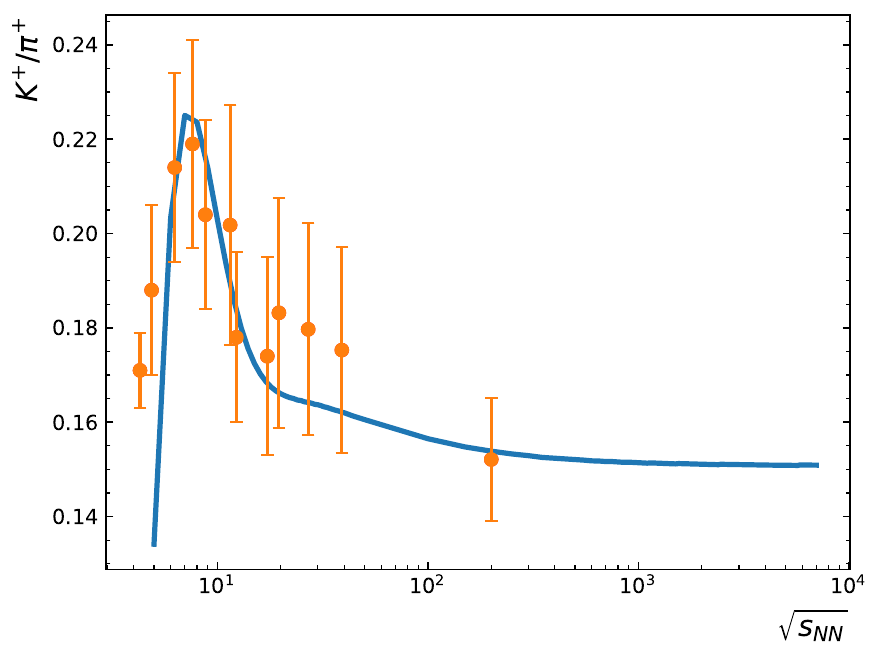}
    \caption{$K^{+}/\pi^{+}$ dependence on $\sqrt{s_{NN}}$}
    \label{fig:kaonpip}
\end{figure}

\begin{figure}
    \includegraphics[scale=0.55]{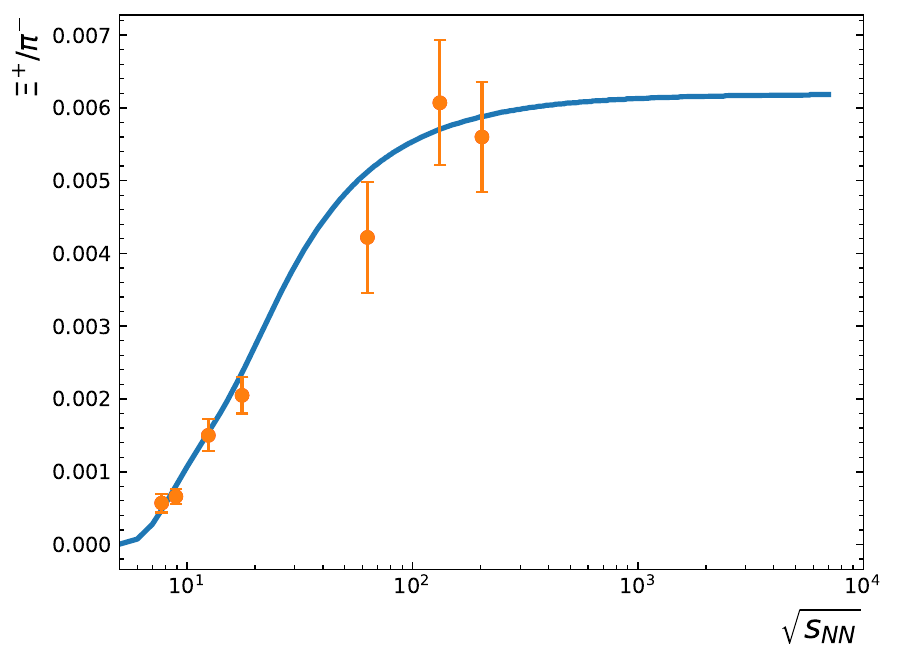}
    \caption{$\Xi^{+}/\pi^{-}$ dependence on $\sqrt{s_{NN}}$}
    \label{fig:cascadeppim}
\end{figure}

With $\gamma_{s}^{o}$ $=$ 0.75, $y$ $=$ 0.34 GeV$^{-1}$ and $z$ $=$ 11.5 GeV. The curve showing the $\sqrt{s_{NN}}$ dependence of $\gamma_{s}$ is shown in Fig.~\ref{fig:gammas}. We find that the factor $\gamma_{s}$ is more suppressed towards the higher collision energies. Using this approach, we redefine the theoretically obtained values of $K^{-}/\pi^{-}$, $K^{+}/\pi^{+}$, $\Xi^{+}/\pi^{-}$ as $\gamma_{s}(K^{-}/\pi^{-})$, $\gamma_{s}(K^{+}/\pi^{+})$, $\gamma_{s}^{2}(\Xi^{+}/\pi^{-})$.
In this approach a factor of $\gamma_{s}$ is multiplied for each strange quark (antiquark) contained in a given hadronic specie in the system. 
We have demonstrated the impact of strangeness suppression factor on the $K^{-}/\pi^{-}$ ratio in Fig.~\ref{fig:kaonpim}. The green line represents a curve without $\gamma_{s}$, while the blue line depicts the results after taking into account this factor i.e., $\gamma_{s}$. It is evident that the experimental data only aligns with the model results after considering this factor. Hence we have applied this factor in all strange-to-nonstrange ratios in this calculation.
The theoretical results of these ratios thus obtained after this \textit{correction} emanating as a consequence of the partial saturation of the strange phase space are shown in Figs.~\ref{fig:kaonpip} and~\ref{fig:cascadeppim}.
\begin{figure}
    \includegraphics[scale=0.55]{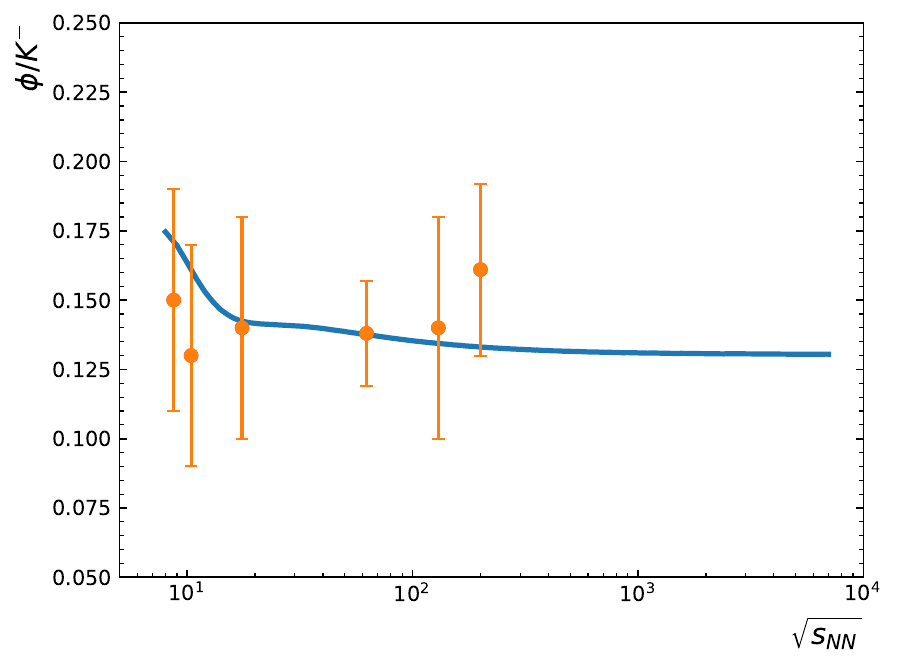}
    \caption{$\phi/K^{-}$ dependence on $\sqrt{s_{NN}}$}
    \label{fig:phikm}
\end{figure}
We find that the theoretical curves after incorporating the effect of strangeness suppression factor fit the data quite reasonably, with $\chi^{2}$/dof values for the three cases being 0.11, 1.20 and 0.85 respectively. 
It is noteworthy that in some previous works~\cite{hp13,hp63,hp67,hp72,hp12,hp82,hpref12,refree2} the factor $\gamma_{s}$ is found to increase with $\sqrt{s_{NN}}$ while in the present case we find that it decreases with $\sqrt{s_{NN}}$. 
This is because of the finite size effects that we include in this calculation while in previous works mentioned above have not incorporated this effect.
The ansatz in Eq.~\eqref{eqn:gammas} takes this aspect into account. This difference seems to emerge from the fact that in the previous analysis the hadrons were treated as purely point-like particles while in our case we have taken into account their finite sizes resulting from the baryonic hard-core repulsion. It thus indicates that the consideration of the baryonic (antibaryonic) finite sizes, leading to excluded volume type effect, play an important role in determining the values of the parameters of thermal models used for analyzing the systems produced in URHIC. The treatment of the hard-core radii of the baryons and antibaryons as a free parameter helps in understanding this aspect of the HRG.

It is interesting to see that the theoretical curve thus obtained by taking into account the finite sizes of baryons and antibaryons as well as the strangeness suppression effect for the case of the $K^{+}/\pi^{+}$ ratio also explains the horn structure in a very reasonable manner. We see that the $K^{-}/\pi^{-}$ and $K^{+}/\pi^{+}$ ratios approach almost the same value $\sim$ 0.15 for large $\sqrt{s_{NN}}$. This again seems to emerge from an almost equal production of $K^{-}$ and $K^{+}$ at the highest RHIC and LHC energies, as indicated in Fig.~\ref{fig:kaons} also and the reaction channels responsible for their productions. We again note from Fig.~\ref{fig:cpottemp} that with an increasing $\sqrt{s_{NN}}$, the temperature $(T)$ of the system increases while the BCP $\mu_{B}$ decreases. Within the framework of the thermal model, both of these effects will lead to an enhanced production of $K^{-}$ towards higher energies. On the other hand, the $K^{+}$ production will be supported by the increasing temperature $(T)$ but the decreasing BCP $(\mu_{B})$ will tend to decrease its production. Due to these two competing effects for $K^{+}$, we find that there is a rapid initial enhancement in the production of $K^{+}$ with increasing $\sqrt{s_{NN}}$, but this enhancement tends to become weaker at larger values of $\sqrt{s_{NN}}$. While the production of pions ($\pi^{-}$ as well as $\pi^{+}$) increases monotonically with increasing temperature $(T)$. This can therefore lead to a horn-like structure in the $K^{+}/\pi^{+}$ variation as shown in Fig.~\ref{fig:kaonpip} and a smooth monotonic rise in the $K^{-}/\pi^{-}$ ratio in Fig.~\ref{fig:kaonpim}. However, one finds that in actual experiments the non-saturation of the strangeness phase space is also found to play an important role in describing the dependence of the ratios $K^{+}/\pi^{+}$ and $K^{-}/\pi^{-}$ on $\sqrt{s_{NN}}$.

In Fig.~\ref{fig:cascadeppim}, we have shown the $\sqrt{s_{NN}}$ dependence of the $\Xi^{+}/\pi^{-}$ ratio. It indicates that the doubly strange antihyperon $(\Xi^{+})$ abundance increases with increasing $\sqrt{s_{NN}}$ rapidly than $\pi^{-}$. This may be understood by noting that though the decreasing value of $\gamma_{s}$ with $\sqrt{s_{NN}}$ tends to lower the $(\Xi^{+})$ abundance but correspondingly the decreasing BCP and increasing temperature tend to enhance its abundance.
In Fig.~\ref{fig:phikm}, we have also shown our results for the $\phi/K^{-}$ ratio variation with the centre of mass collision energy in comparison with the experimental data~\cite{hpref8}. A reasonably good fit is obtained again by considering a partial saturation of the strange sector. A relatively rapid decrease in the $K^{-}$ abundance 
due to rapidly increasing $(\mu_{B})$ and $\gamma_{s}$ towards lower energies is seen to give rise in the $\phi/K^{-}$ for smaller $\sqrt{s_{NN}}$.

\section{Summary and Conclusion} 
\label{summary}
We have attempted to analyze the energy dependence of several particle ratios over a wide range of collision energies in URHIC. The available experimental data are described quite well in the framework of a statistical thermal model, which also incorporates an important feature of the baryonic (antibaryonic) interaction, namely the hard-core repulsion, leading to an excluded volume type effect among the baryons (antibaryons). It is treated as a free model parameter and is found that it has a significant effect on the relative abundance of hadrons produced in such collisions. A hard-core radius of 0.76 to 0.79 fm is required to obtain a good fit to the experimental data over the wide range of energy considered. Invoking the HRG model we fix the values of the freeze-out BCP and $T$ at different collision energies using the standard ansatzes. Contributions of weak decays of the heavy hadronic resonance up to the 2 GeV mass to the final state relative hadron multiplicities are taken into account. For this purpose single weak decays and those double decays where weak decay is followed by the strong decay have been taken into account. Minimum $\chi^{2}$ fits to theoretically describe the variation of the experimental particle ratios with collision energy are obtained by treating baryon and antibaryon hard-core radius $r_{B}$ as a free parameter for each type of antibaryon-to-baryon ratio. We also find indications of two different freeze-out stages. The earlier one corresponds to the baryonic (hyperonic) and antibaryonic (antihyperonic) degrees of freedom and the later stage corresponds to freeze-out of the mesonic degrees of freedom such as $K$ and $\pi$. However, the $\phi$ meson due its small reaction cross section with other hadrons is assumed to freeze-out earlier than $K$ and $\pi$. The abundance of strange hadrons relative to pions is found to be suppressed in the experiments as compared to the theoretical predictions. This lends support to the idea of the strangeness suppression factor $(\gamma_{s})$. Using the experimental data, we have also extracted a suitable ansatz to define $(\gamma_{s})$ over the wide range of collision energy. The $\gamma_{s}$ is found to decrease in our case while in some earlier works it is shown to increase with $\sqrt{s_{NN}}$. This contrast seems to emerge from the fact that in the previous analysis the hadrons were treated as point-like particles. Our results indicate a high degree of relative strangeness suppression at larger collision energies in the experiments compared to the pions which are considered to be a major indicator of the entropy content of the system. The horn structure seen in the energy dependence of the $K^{+}/{\pi^{+}}$ ratio is reasonably described.

\bibliographystyle{apsrev4-2}
\bibliography{hyperon_production}
\end{document}